\documentstyle[preprint,prc,aps,eqsecnum,epsfig]{revtex}
\everymath={\displaystyle}
\begin{document}
\title{\bf Nucleon-nucleon scattering in a chiral constituent quark model}
\author{D. Bartz\footnote{e-mail : d.bartz@ulg.ac.be} and
Fl. Stancu\footnote{e-mail: fstancu@ulg.ac.be}}

\address{Universit\'e de
Li\`ege, Institut de Physique B5, Sart Tilman, B-4000 Li\`ege~1, Belgium}
\date{\today}
\maketitle

\begin{center}
\begin{minipage}{0.99\textwidth}
\begin{abstract}
\baselineskip=0.50cm
We study the nucleon-nucleon interaction in a chiral constituent quark model by
using the resonating group method, convenient for treating the interaction
between composite particles. The calculated phase shifts for the $^3S_1$ and
$^1S_0$ channels show the presence of a strong repulsive core due to the
combined effect of the quark interchange and the spin-flavour structure of the
effective quark-quark interaction. Such a symmetry structure stems from the
pseudoscalar meson exchange between the quarks and is a consequence of the 
spontaneous breaking of the chiral symmetry. We perform single and coupled
channel calculations and show the role of coupling of the $\Delta\Delta$ and
hidden color $CC$ channels on the behaviour of the phase shifts.
\end{abstract}
\pacs {PACS numbers: }
\end{minipage}
\end{center}


\thispagestyle{empty}
\section {Introduction}
Many studies have been devoted so far to the understanding of the
nucleon-nucleon ($NN$) interaction starting from models which have been
considered to be successful in baryon spectroscopy. Here we refer to
nonrelativisitc quark models in the framework of which calculations of
scattering phase shifts can be made quantitatively. We can roughly divide
these models into three categories. In the first category we consider models
based on one-gluon exchange (OGE) between quarks. They explain the short-range
repulsion of the $NN$ potential as due to the chromomagnetic spin-spin part
of OGE, combined with quark interchanges between $3q$ clusters (for a review
see e.g. \cite{OKA85,MYH88,SHI89,STB00}). In addition, the long-range part is
obtained from the one-pion exchange potential acting directly between two
nucleons and the medium-range part is introduced phenomenologically as a local
central potential \cite{OKA83}.

There is a second category, of hybrid models \cite{KUS91,ZHA94,FUJ96}, where in
addition to OGE, quarks belonging to different clusters interact also via
pseudoscalar and scalar meson exchange. In these hybrid models the short-range
repulsion is still attributed to OGE and the middle- and long-range attraction
is due to meson exchanges between quarks.

Here we employ a model of the third category where the quark-quark interaction,
besides the confinement, is due entirely to meson exchanges between quarks.
This is the chiral constituent quark model proposed in Ref. \cite{GLO96a} and
parametrized in a nonrelativistic version in Refs. \cite{GLO96b,GLO97}. There
are also semirelativistic versions available, see e. g. \cite{GLO98}. For the
present status of the model we refer the reader to Ref. \cite{GLO00}.

The origin of the model \cite{GLO96a,GLO96b,GLO97,GLO98,GLO00} is thought to
lie in the spontaneous breaking of chiral symmetry in QCD which implies the
existence of constituent quarks with a dynamical mass and Goldstone bosons
(pseudoscalar mesons). According to the two-scale picture of Manohar and
Georgi \cite{MAN94} at a distance beyond that of spontaneous chiral symmetry
breaking, but within that of the confinement scale, the appropriate degrees of
freedom should be the constituent quarks and the chiral meson fields. If a
quark-pseudoscalar meson coupling is assumed, in a nonrelativistic limit one
obtains a quark-meson vertex proportional to
$\vec{\sigma}\cdot\vec{q}~\lambda^F$ with $\vec{\sigma}$ the Pauli matrices,
$\vec{q}$ the momentum of the meson and $\lambda^F$ the Gell-Mann flavor ($F$)
matrices. This generates a pseudoscalar meson exchange interaction between
quarks which is spin and flavor dependent.

In the following this interaction is referred to as a Goldstone boson exchange
(GBE) interaction. In the coordinate space the corresponding potential contains
two terms. One is a repulsive Yukawa potential tail and the other is an
attractive contact $\delta$-interaction. When regularized \cite{GLO96b,GLO97}
the latter generates the short range part of the quark-quark interaction. The
short-range part dominates over the Yukawa part in the description of baryon
spectra leading to a correct order of positive and negative parity states
\cite{ITW88}. This interaction contains the main ingredients required in the
calculation of the $NN$ potential, and it is thus natural to study the $NN$
problem within this model. In addition, the two-pion exchange interaction
between constituent quarks reinforces the effect of the flavor-spin part of
the one-meson exchange and also provides a contribution of a $\sigma$-type
scalar meson exchange \cite{RIS99} required to describe the middle-range
attraction.

The spin-flavor symmetry structure of the model
\cite{GLO96a,GLO96b,GLO97,GLO98,GLO00}, which is essential in describing the
light baryon spectrum is getting support from the phenomenological analysis of
$L=1$ negative parity resonances \cite{COL99}. Also $1/N_c$ QCD studies
\cite{CAR00} have a consistent interpretation in a constituent quark model
with pseudoscalar meson exchange interaction. The spontaneous
chiral symmetry breaking is responsible for the absence of parity doubling in
low energy hadron spectrum. In particular it explains the splitting between the
negative parity state $N^*$(1535) and the nucleon $N$(939). The quark models,
as e. g. the OGE model, which explicitly breaks chiral symmetry, fail to
reproduce the $N^*(1535)-N(939)$ splitting. Recent lattice calculations,
which take into account the chiral symmetry of QCD
\cite{SAS00}, were able to reproduce the above $N^*-N$ splitting. This brings new
substantial support to the model \cite{GLO96a,GLO96b,GLO97,GLO98,GLO00}.

This work is a natural extension of the previous studies
\cite{STA97,BAR99a,BAR99b}. Ref. \cite{STA97} was rather exploratory about the
role of a spin-flavor dependent interaction in giving rise to a repulsive core.
Within the parametrization \cite{GLO96b} of the GBE model it was found that at
zero-separation between two $3q$ clusters the height of the repulsive core is
0.830 GeV and 1.356 GeV in the $^3S_1$ and $^1S_0$ channels respectively. The
spin-flavor symmetry and the parametrization \cite{GLO96b} of the GBE model
favours the $|\left[42\right]_{O}\left[51\right]_{FS}\rangle$ state which
becomes highly dominant. In a better basis \cite{BAR99a}, obtained from
single-particle molecular type states, instead of cluster model states, the
situation is similar, the repulsion being reduced by about 200 MeV in the
$^3S_1$ channel and by about 400 MeV in the $^1S_0$ channel. This is natural
because the molecular orbital basis gives a lower bound of the expectation
value of the Hamiltonian in the six-quark basis. In Ref. \cite{BAR99b} an
adiabatic nucleon-nucleon potential was calculated based on the model
\cite{GLO96b}. It was found that none of the bases, cluster or molecular,
leads to an attractive pocket. An attraction was simulated by introducing a
$\sigma$-meson exchange of a similar analytic structure between quarks, with
that of the pseudoscalar meson exchange.

Here, instead of \cite{GLO96b}, we use the chiral
constituent quark model version of Ref. \cite{GLO97} where the GBE interaction
is parametrized in a more realistic way. The adiabatic potential calculated
\cite{BAR99b}  in the Born-Oppenheimer approximation with this version posesses
a small attractive pocket, in contrast to that resulting from model
\cite{GLO96b} (see Ref. \cite{BAR99a}).

The present study is based on a dynamical approach to the $NN$ interaction,
namely the resonating group method (RGM) \cite{WHE37,WIL77,KAM77}, which allows
the calculation of both bound states and phase shifts. This method has already
been used in $NN$ studies with models of categories 1) and 2) mentioned above.
So far it has been always applied to nonrelativistic models, where the wave function of 
the nucleon can be approximated by an $s^3$ configuration.

In the next section we shortly review the Hamiltonian model \cite{GLO97}. In
Sec. III we describe the main steps of the resonating group method for bound
and scattering states. The $6q$ basis formed of $NN$, $\Delta\Delta$ and $CC$
(hidden color) states is introduced in subsection III~C. In Sec. IV we derive
the matrix elements required by the RGM method for the typical spin-flavor
structure of the GBE model. In Sec. V we present the results for the phase
shifts in the $^3S_1$ and $^1S_0$ channels and discuss the role of the coupled
$\Delta\Delta$ and $CC$ channels on the $NN$ phase shifts. The last section is
devoted to conclusions.

\section{The model}

The GBE Hamiltonian considered below has the form \cite{GLO97}
\begin{equation}\label{HAMILTONIAN}
H=\sum_i m_i+\sum_{i=1}{\frac{p^2_i}{2m_i}}-K_G+\sum_{i<j} V_{Conf}(r_{ij})+\sum_{i<j} V_{\chi}(r_{ij})\ ,
\end{equation}
where $K_G$ is the kinetic energy of the center of mass. The linear confining
interaction is
\begin{equation}
V_{Conf}(r_{ij})=-\frac{3}{8}\lambda^c_i \cdot \lambda^c_j \ (C r_{ij} + V_0)\ ,
\end{equation}
and the spin-spin component of the GBE interaction in its $SU_F(3)$ form is
\begin{equation}\label{CHIRAL}
V_{\chi}(r_{ij})=\{ {\sum_{F=1}^3 V_{\pi}(r_{ij}) \lambda^F_i \lambda^F_j + \sum_{F=4}^7 V_K(r_{ij}) \lambda^F_i \lambda^F_j + V_{\eta}(r_{ij}) \lambda^8_i \lambda^8_j + \frac{2}{3} V_{\eta'}(r_{ij}) } \} \vec{\sigma_i} \cdot \vec{\sigma_j}\ .
\end{equation}
The interaction (\ref{CHIRAL}) contains $\gamma=\pi,K,\eta,$ and $\eta'$
meson exchange terms and $V_{\gamma}(r_{ij})$ is given as the sum of two
distinct contributions: a Yukawa-type potential containing the mass of the
exchanged meson and a short-range contribution of opposite sign, the role of
which is crucial in baryon spectroscopy. For a given meson $\gamma$, the
exchange potential is
\begin{equation}\label{GRAZ}
V_{\gamma}(r) = \frac{g_{\gamma}^2}{4\pi} \frac{1}{12m_im_j} \{ \mu_{\gamma}^2 \frac{e^{- \mu_{\gamma} r}}{r} - \Lambda_{\gamma}^2 \frac{e^{- \Lambda_{\gamma} r}}{r} \} \ ,
\end{equation}
where $\Lambda_{\gamma}=\Lambda_0+\kappa \mu_{\gamma}$. For a system of $u$
and $d$ quarks only, as is the case here, the $K$ exchange does not contribute.
In the calculations below we use the parameters of Ref. \cite{GLO97}. These are
$$m_{u,d}=340\ {\rm MeV}\ ,\ \ C=0.77\ {\rm fm}^{-2}\ ,$$
$$\mu_{\pi}=139\ {\rm MeV}\ ,\ \mu_{\eta}=547\ {\rm MeV}\ ,\ \mu_{\eta'}=958\ {\rm MeV}\ ,$$
$$\frac{g_{\pi q}^2}{4\pi} = \frac{g_{\eta q}^2}{4\pi} = 1.24\ ,\ \ \frac{g_{\eta' q}^2}{4\pi} = 2.7652\ ,$$
\begin{equation}\label{PARAM}
\ \ \ \Lambda_0=5.82\ {\rm fm}^{-1}\ ,\  \kappa = 1.34\ ,\  V_0=-112\ {\rm MeV}\ .
\end{equation}

As already mentioned before, the reason of using the parametrization
\cite{GLO97}, instead of \cite{GLO96b}, as in the previous work
\cite{STA97,BAR99a,BAR99b}, is that it is more realistic. Its volume integral,
i. e. its Fourier transform at $\vec{q} = 0$, vanishes, consistently with the
quark-pseudoscalar meson vertex proportional to
$\vec{\sigma} \cdot \vec{q}\ \lambda^F$. In addition, this interaction does not
enhance the quark-quark matrix elements containing $1p$ relative motion, as it
is the case with the parametrization \cite{GLO96b}. This point has been raised
in Ref. \cite{STA99}.

At this stage we wish to stress that the above parametrization gives a good
description of baryon spectra. We do not change any parameter obtained from the
fit \cite{GLO97}. Such a parametrization is, of course, only effective.
However, irrespective of the parametrization, the flavor-spin symmetry is
essential in this model. There are also semirelativistic versions of the GBE
model, as for example \cite{GLO98} but the application of the RGM techniques to
semirelativistic six-quark Hamiltonians is certainly much more involved.

\section{The resonating group method}
The resonating group method \cite{WHE37} is one of the well established methods
used to study the interaction between two composite systems. It allows to
calculate bound states energies and scattering phase shifts. It has been first
applied to nuclear physics in the study of the nucleus-nucleus interaction
\cite{WIL77,KAM77}. Its application to baryon-baryon systems was initiated by
Oka and Yazaki \cite{OKA80}. In a baryon-baryon system, where each baryon is a
$3q$ cluster, it takes explicitly into account the quark interchange between
the two interacting baryons. This comes from the assumption that the total wave
function can be written as
\begin{equation}\label{PSYCC}
\psi=\sum\limits_{\beta}^{}{\mathcal{A}}\left[\Phi_{\beta}
\chi_{\beta}(\vec{R}_{AB})\right]\ ,
\end{equation}
where $\beta$ is a specific channel (here $\beta=NN$, $\Delta\Delta$ or $CC$),
${\mathcal{A}}$ is an antisymmetrization operator defined below, $\Phi_{\beta}$
contains the product of internal wave functions of the interacting baryons and
$\chi_{\beta}(\vec{R}_{AB})$ is the wave function of the relative motion in the
channel $\beta$, depending on the relative coordinate $\vec{R}_{AB}$ between
clusters $A$ and $B$.

The internal wave function of each cluster has orbital, flavor, spin and color
parts. In $\Phi_{\beta}$ the flavor and spin are combined to give a definite
total spin $S$ and isospin $I$ so that one has
\begin{equation}\label{PHI}
\Phi_{\beta}=\left[\phi_A(\vec{\xi}_A)\phi_B(\vec{\xi)}_B)\right]_{SI}\ ,
\end{equation}
where $\vec{\xi}_A = (\vec{\xi}_1,\vec{\xi}_2)$ and
$\vec{\xi}_B = (\vec{\xi}_3,\vec{\xi}_4)$ are the internal coordinates of the
clusters $A$ and $B$:
\begin{eqnarray}\label{JACOBI}
\vec{\xi}_1 = \vec{r}_1 - \vec{r}_2\ ,\ \ &&\ \ \vec{\xi}_3 = \vec{r}_4 - \vec{r}_5\ ,\nonumber \\
\vec{\xi}_2 = \frac{\vec{r}_1 + \vec{r}_2 - 2 \vec{r}_3}{2}\ ,\ \ && \ \ \vec{\xi}_4 = \frac{\vec{r}_4 + \vec{r}_5 - 2 \vec{r}_6}{2}\ ,\nonumber \\
\vec{R}_A   = \frac{\vec{r}_1 + \vec{r}_2 + \vec{r}_3}{3}\ ,\ \ &&\ \ \vec{R}_B = \frac{\vec{r}_4 + \vec{r}_5 + \vec{r}_6}{3}\ .
\end{eqnarray}
The functions $\phi_i(\xi_i),\ i=A,B$ are supposed to be known (see later). 
They are totally antisymmetric $3q$ states in orbital, spin, flavor and color
space. The color part is a $[1^3]$ singlet for $N$ and $\Delta$ states and an
octet for $C$ states. Usually the color part of a $3q$ state is not written
explicitly. The same statement remains valid for the $6q$ state which is a
$[222]_C$ singlet in any channel. 

The antisymmetrization operator ${\mathcal{A}}$ is defined by
\begin{equation}
{\mathcal{A}}=1-\sum\limits_{i=1}^{3}\sum\limits_{j=4}^{6}P_{ij}\ ,
\end{equation}
where $P_{ij}$ is the permutation operator of the quarks $i$ and $j$ belonging
to clusters $A(1,2,3)$ and $B(4,5,6)$ respectively. It acts in the orbital, 
flavor, spin and color space, so it can be written as
$P_{ij}=P_{ij}^oP_{ij}^fP_{ij}^{\sigma}P_{ij}^c$ where
\begin{equation}\label{PIJ}
P_{ij}^f=\frac{1}{2}\lambda_i^f \cdot \lambda_j^f+\frac{1}{3}\ ,~\ \ \ \ 
P_{ij}^{\sigma}=\frac{1}{2}\vec{\sigma}_i \cdot \vec{\sigma}_j+\frac{1}{2}\ ,~\ \ \ \ 
P_{ij}^c=\frac{1}{2}\lambda_i^c \cdot \lambda_j^c+\frac{1}{3}\ ,
\end{equation}
with $\lambda_i^{f(c)}$ the Gell-Mann matrices of $SU_F(3)$ ($SU_C(3)$) and
$\vec\sigma_i$ the Pauli matrices.

Let us first consider the one channel case. From the variational principle one
can obtain the equation determining the relative wave function
$\chi(\vec{R}_{AB})$
\begin{equation}\label{RGM1}
\int \phi^+(\vec{\xi}_A)\phi^+(\vec{\xi}_B)(H-E)
{\cal A} [\phi(\vec{\xi}_A)\phi(\vec{\xi}_B)\chi(\vec{R}_{AB})] d^3\xi_A d^3\xi_B = 0\ ,
\end{equation}
where $H$ is the Hamiltonian of the six-quark system. As usually (see e. g. 
Ref. \cite{SHI89}) we introduce the Hamiltonian kernel
\begin{eqnarray}\label{HKERNEL}
{\cal H}(\vec{R'},\vec{R})&=&\int \phi^+(\vec{\xi}_A)\phi^+(\vec{\xi}_B) 
\delta(\vec{R'}-\vec{R}_{AB}) H {\cal A} [\phi(\vec{\xi}_A)\phi(\vec{\xi}_B) 
\delta(\vec{R}-\vec{R}_{AB})] d^3\xi_A d^3\xi_B d^3R_{AB} \nonumber \\
&=& {\cal H}^{(d)}(\vec{R})\delta(\vec{R}-\vec{R'})-{\cal H}^{(ex)}(\vec{R'},\vec{R})\ ,
\end{eqnarray}
and the normalization kernel
\begin{eqnarray}\label{NKERNEL}
{\cal N}(\vec{R'},\vec{R})&=&\int \phi^+(\vec{\xi}_A)\phi^+(\vec{\xi}_B) 
\delta(\vec{R'}-\vec{R}_{AB}) {\cal A} [\phi(\vec{\xi}_A)\phi(\vec{\xi}_B) 
\delta(\vec{R}-\vec{R}_{AB})]d^3\xi_A d^3\xi_B d^3R_{AB} \nonumber \\
&=& {\cal N}^{(d)}(\vec{R})\delta(\vec{R}-\vec{R'})-{\cal N}^{(ex)}(\vec{R'},\vec{R})\ .
\end{eqnarray}
The direct term of the Hamiltonian kernel, ${\cal H}^{(d)}(\vec{R})$, consists of the relative kinetic, the relative potential and the internal energies :
\begin{equation}\label{HDIRECT}
{\cal H}^{(d)}(\vec{R})=-\frac{\nabla^2_R}{2\mu}+V_{rel}^{(d)}(\vec{R})+H_{int}\ ,
\end{equation}
where $\mu=3m/2$ is the reduced mass of the clusters $A$ and $B$. Then Eq.
(\ref{RGM1}) can be written as
\begin{equation}
\int {\cal L}(\vec{R'},\vec{R}) \chi(\vec{R}) d^3R = 0\ ,
\end{equation}
where ${\cal L}(\vec{R'},\vec{R}) = {\cal H}(\vec{R'},\vec{R}) - E {\cal N}(\vec{R'},\vec{R})$. This is the RGM equation. Using (\ref{HDIRECT}) one can write
\begin{equation}
{\cal L}(\vec{R'},\vec{R}) = [-\frac{\nabla^2_R}{2\mu}+V_{rel}^{(d)}(\vec{R})-E_{rel}]\delta(\vec{R}-\vec{R'})-[{\cal H}^{(ex)}(\vec{R'},\vec{R})-E{\cal N}^{(ex)}(\vec{R'},\vec{R})]\ ,
\end{equation}
where $E_{rel}=E-H_{int}$ is the energy of the relative motion. There are two
important steps in solving this equation. One is to calculate the Hamiltonian
kernel (\ref{HKERNEL}) by reducing the six-body matrix elements to two-body
matrix elements. This is discussed in Sec IV. Another step is the
discretisation of the RGM equation. It is important both for bound and
scattering states. The discretisation has been performed by using the method
of Ref. \cite{KAM77}.
\subsection{Bound states}
Here we briefly describe the discretisation procedure directly applicable to
bound states. According to Ref. \cite{KAM77}, the relative wave function
$\chi (\vec{R})$ has been expanded over a finite number of 
Gaussians $\chi_i$ centered at $\vec{R}_i$ ($i = 1,2,...,N$) where $R_i$ are points, 
here equally spaced,
between the origin and some value of $R$ depending on the range of the
interaction. The expansion is
\begin{equation}\label{KAM1}
\chi(\vec{R})=\sum_{i=1}^N C_i \chi_i(\vec{R})\ ,
\end{equation}
with
\begin{equation}\label{KAM1b}
\chi_i(\vec{R})=g(\vec{R}-\vec{R_i},\sqrt{2/3 b})=(\frac{3}{2 \pi b^2})^{3/4} e^{-\frac{3}{4 b^2}(\vec{R}-\vec{R}_i)^2}\ . 
\end{equation}
If $g(\vec{r},b)$ is the normalized Gaussian wave function of a quark,
given by
\begin{equation}\label{QUARKGAUSS}
g(\vec{r},b)=(\frac{1}{\pi b^2})^{3/4} e^{-\frac{r^2}{2 b^2}}\ , 
\end{equation}
from the Jacobi transformations (\ref{JACOBI}) it follows that the relative wave 
function is expanded in terms of the
Gaussians (\ref{KAM1b}) with the size parameter $\sqrt{2/3 b}$. This
method can be applied straightforwardly to the bound state problem. The
modification necessary for treating the scattering problem will be explained
later in the next subsection. The binding energy $E$ and the expansion
coefficients $C_i$ are given by the eigenvalues and eigenvectors of the
following equation :
\begin{equation}\label{GCM}
\sum_{j=1}^N H_{ij}C_j = E \sum_{j=1}^N N_{ij}C_j\ ,
\end{equation}
where $N$ is the number of Gaussians considered in (\ref{KAM1}). The matrices 
\begin{equation}
H_{ij}=\int \phi^+(\vec{\xi}_A)\phi^+(\vec{\xi}_B) 
\chi(\vec{R}_{AB}-\vec{R}_i) H (1-{\cal A}') [\phi(\vec{\xi}_A)\phi(\vec{\xi}_B) 
\chi(\vec{R}_{AB}-\vec{R}_j)] d^3\xi_A d^3\xi_B d^3R_{AB}\ ,
\end{equation}
and
\begin{equation}
N_{ij}=\int \phi^+(\vec{\xi}_A)\phi^+(\vec{\xi}_B) 
\chi(\vec{R}_{AB}-\vec{R}_i) (1-{\cal A}') [\phi(\vec{\xi}_A)\phi(\vec{\xi}_B) 
\chi(\vec{R}_{AB}-\vec{R}_j)] d^3\xi_A d^3\xi_B d^3R_{AB}
\end{equation}
are obtained from (\ref{HKERNEL}) and (\ref{NKERNEL}) respectively. By
including the center of mass coordinate $(\vec{R}_A+\vec{R}_B)/2$ and transforming 
back to
$r_i\ (i=1,...,6)$ we get the following formulas
\begin{equation}
H_{ij}=\int \prod_{k=1}^3 \phi^+(\vec{r}_k-\frac{\vec{R}_i}{2}) \prod_{k'=4}^6 \phi^+(\vec{r}_{k'}+\frac{\vec{R}_i}{2}) H {\cal A} [\prod_{l=1}^3\phi(\vec{r}_l-\frac{\vec{R}_j}{2}) \prod_{l'=4}^6\phi(\vec{r}_{l'}+\frac{\vec{R}_j}{2})] d^3r_1 ...  d^3r_6\ ,
\end{equation}
and
\begin{equation}
N_{ij}=\int \prod_{k=1}^3 \phi^+(\vec{r}_k-\frac{\vec{R}_i}{2}) \prod_{k'=4}^6 \phi^+(\vec{r}_{k'}+\frac{\vec{R}_i}{2}) {\cal A} [\prod_{l=1}^3\phi(\vec{r}_l-\frac{\vec{R}_j}{2}) \prod_{l'=4}^6\phi(\vec{r}_{l'}+\frac{\vec{R}_j}{2})] d^3r_1 ...  d^3r_6\ ,
\end{equation}
with $\phi(\vec{r})=g(\vec{r},b)$ given by (\ref{QUARKGAUSS}).
These forms are much easier to handle in actual calculations. They allow to
reduce the $6q$ matrix elements to two-body matrix elements. Moreover the
distances $R_i$ play now the role of a generator coordinate \cite{STB00} and
lead to a better understanding of the relation between the resonating group
method and the generator coordinate method \cite{CVE83}.
\subsection{Scattering states}
For scattering states the expansion (\ref{KAM1}) holds up to a finite distance
$R=R_c$, depending on the range of the interaction. Beyond $R_c$,
$\chi(\vec{R})$ becomes the usual combination of Hankel functions containing
the $S$-matrix. Because practical calculations of both bound state and
scattering states are done in terms of partial waves, we first give the partial
wave expansion of Eq. (\ref{KAM1}) in terms of locally peaked wave functions
with a definite angular momentum $l$ and projection $m$:
\begin{equation}\label{KAM3}
\chi_{lm}(\vec{R})=\sum_{i=1}^N C_i^{(l)} \chi_i^{(l)}(R)Y_{lm}(\hat{R})\ ,
\end{equation}
with the explicit form of $\chi_i^{(l)}$ given by
\begin{equation}\label{KAM4}
\chi_i^{(l)}(R)=4 \pi (\frac{3}{2 \pi b^2})^{3/4} e^{-\frac{3}{4 b^2}(R^2+R_i^2)}i_l(\frac{3}{2 b^2}R R_i)\ ,
\end{equation}
where $i_l$ is the modified spherical Bessel function \cite{ABR64}. When we
treat the scattering problem, the form (\ref{KAM4}) holds up to $R\leq R_c$
only. In fact in this case the relative wave function is expanded in terms of
${\tilde{\chi}}^{(l)}$ as
\begin{equation}\label{KAM5}
\chi^{(l)}(R)=\sum_{i=1}^N C_i^{(l)}{\tilde{\chi}}_i^{(l)}(R)\ ,
\end{equation}
where
\begin{eqnarray}
{\tilde{\chi}}_i^{(l)}(R) &=& \alpha_i^{(l)} \chi_i^{(l)} (R)\ ,\ \ \ \ \ \ \ \ \ \ \ \ \ \ (R \leq R_c) \nonumber \\
{\tilde{\chi}}_i^{(l)}(R) &=& h_l^{(-)}(kR) + S_i^{(l)}  h_l^{(+)} (kR)\ ,\ \ (R \geq R_c)
\end{eqnarray}
with $\chi_i^{(l)}(R)$ defined by Eq. (\ref{KAM4}). Here $k$ is the wave number
$k=\sqrt{2\mu E_{rel}}$ and $h_l^{(-)}$ and $h_l^{(+)}$ are spherical Hankel
functions \cite{ABR64}. The coefficients $\alpha_i^{(l)}$ and $S_i^{(l)}$ are
determined from the continuity of ${\tilde{\chi}}_i^{(l)}$ and its derivative
at $R=R_c$. The coefficients $C_i^{(l)}$ of (\ref{KAM3}) are normalized such
that $\sum_{i=1}^N C_i^{(l)} = 1$. Then the $S$-matrix is given in terms of the
coefficients $C_i^{(l)}$ as
\begin{equation}\label{KAM6}
S^{(l)}=\sum_{i=1}^N C_i^{(l)} S_i^{(l)}\ . 
\end{equation}
The method of determining the expansion coefficients is described in detail by
Oka and Yazaki \cite{OKA80}.
\subsection{Coupled channels}
Here we consider more than one channel. In this case, based on Eq.
(\ref{PSYCC}), the RGM equation becomes a system of coupled channel equations
for $\chi_{\beta}$
\begin{equation}
\sum_{\beta}\int{\cal L}_{\alpha\beta}(\vec{R'},\vec{R}) \chi_{\beta}(\vec{R}) d^3R = \sum_{\beta}\int [{\cal H}_{\alpha\beta}(\vec{R'},\vec{R})-E {\cal N}_{\alpha\beta}(\vec{R'},\vec{R})] \chi_{\beta}(\vec{R}) d^3R =0\ .
\end{equation}
Usually the normalisation kernel ${\cal N}_{\alpha\beta}$ is not diagonal
because of the antisymmetrisation. For a given $SI$ sector one can establish
which are the $6q$ states of (\ref{PHI}) allowed by the Pauli principle
\cite{HAR81}. Here we consider the $l = 0$ partial waves i. e. we study the
$^3S_1$ and $^1S_0$ phase shifts. In this case, according to \cite{HAR81}, 
the $6q$ allowed states are $NN, \Delta\Delta$ and $CC$. The $NN$ and
$\Delta\Delta$ states are easy to define directly from Eq. (\ref{PSYCC}). For
$CC$ states we adopt the definition of Ref. \cite{FAE82} which is more
appropriate for RGM calculation. This $CC$ state of six quarks allows some
``color polarisation'' of the $6q$ system in the interaction region. It is
defined in the following way
\begin{equation}
|CC\rangle = \alpha |NN\rangle +\beta |\Delta\Delta\rangle +\gamma {\cal A}_{STC}|\Delta\Delta\rangle\ ,
\end{equation}
with
\begin{equation}
{\cal A}_{STC}=\frac{1}{10}[1-\sum_{i=1}^3\sum_{j=4}^6P_{ij}^{\sigma}P_{ij}^{f}P_{ij}^{c}]\ ,
\end{equation}
where $P_{ij}^{\sigma}$,$P_{ij}^{f}$ and $P_{ij}^{c}$ are the exchange
operators in the spin, isospin and color space respectively defined by
(\ref{PIJ}). From the orthonormality conditions $\langle CC|CC\rangle =1$,
$\langle CC|NN\rangle =0$ and $\langle CC|\Delta\Delta\rangle =0$ one can
determine the coefficients $\alpha$, $\beta$ and $\gamma$ so that
\begin{equation}\label{HIDDEN}
|CC\rangle = -\frac{\sqrt{5}}{6} |NN\rangle +\frac{1}{3} |\Delta\Delta\rangle -\frac{15}{4} {\cal A}_{STC}|\Delta\Delta\rangle\ .
\end{equation}
The important feature in the definition of the $CC$-state is that the
eigenvalue of the color $SU(3)$ Casimir operator is 12 for each $3q$ cluster.
This tells us that $C$ is a color octet state and thus explains why we call
the $CC$-state a hidden color state. Note that at zero separation between
quarks (shell model basis) the $CC$ state above is the same as that introduced
by Harvey. The two differ only at finite separation distances. To see the
identity with Harvey's $CC$ state \cite{HAR81} at zero separation one can
combine it with the $NN$ and $\Delta\Delta$ states as defined by Eq. 
(\ref{PSYCC}) to get symmetry states of the form 
$| [f]_{FS} [222]_C ; {\tilde{g}}_{FSC}\rangle$ where ${\tilde{g}}$ is the
representation resulting from the inner product of $[f]_{FS}$ and $ [222]_C$
which is conjugate  with the symmetry $g$ of an orbital state such as to
produce a totally antisymmetric $6q$ state. Comparing Table 3 of Ref.
\cite{FAE82} with that of Harvey's \cite{HAR81} Table 1 one can see that the
coefficients of this basis transformation are identical which proves the
identity of the hidden color state (\ref{HIDDEN}) with that of Harvey at $R=0$.
Note that Harvey's definition \cite{HAR81} of $CC$ is more appropriate for
generator coordinate method than for RGM calculations.
\section{Six-body matrix elements}

The method to compute the six-body matrix elements is explained in some detail
in the appendix. In Tables I \& II we give the results for diagonal and
off-diagonal matrix elements of the channels $NN$, $\Delta\Delta$ and $CC$ to
be used in coupled channel calculations of the $^3S_1$ and $^1S_0$ phase shifts
respectively. Although we apply the SU(3) version of the GBE model the matrix
elements of $\sigma_{i} \cdot \sigma_{j}~\tau_{i} \cdot \tau_{j}$ and
$\sigma_{i} \cdot \sigma_{j}~\tau_{i} \cdot \tau_{j} P^{f \sigma c}_{36}$ 
needed in SU(2) calculations are also indicated. In fact they are used in
calculating the expectation value of $\sigma_{i} \cdot \sigma_{j}~ \lambda^8_i
\cdot \lambda^8_j$ by subtracting them from
$\sigma_{i} \cdot \sigma_{j}~\lambda^f_i \cdot \lambda^f_j$ because there is
no $K$ meson exchange. Moreover the values we found can be considered as a
validity test of our method because they are in full agreement with Table 1
of Ref. \cite{KS84}.
\section{Numerical results}
We perform RGM calculation as described above for $NN$, $NN+\Delta \Delta$ and
$NN + \Delta \Delta +CC$ channels. In all cases the size parameter of the
Gaussian (\ref{QUARKGAUSS}) is fixed at $b=0.44$ fm by the stability condition
(see for example Ref.\cite{OKA85}) 
\begin{equation}\label{STABILITY}
\frac{\partial}{\partial b}\langle \phi|H|\phi\rangle =0\ ,
\end{equation}
where $\phi$ is a variational solution of the Hamiltonian (\ref{HAMILTONIAN})
for a ground state $3q$ system. This solution is fully symmetric in the orbital
space and is chosen to be of the form
\begin{equation}\label{NUCLEONGROUND}
\phi=\prod_{i=1}^3 g(\vec{r}_i,b)\ ,
\end{equation}
with $g(\vec{r}_i,b)$ of (\ref{QUARKGAUSS}).

Either if we take  one, two or three channels namely $NN$, $NN+\Delta\Delta$ or
$NN+\Delta\Delta+CC$ we found that a number of 15 Gaussians in the expansion
(\ref{KAM1}) is large enough to obtain convergence. In all cases the result is
stable at the matching radius $R_c = 4.5$ fm. In Figs. 1 \& 2 we show the phase
shifts as a function of the relative momentum $k$ obtained from one, two and
three coupled channels. One can see that the addition to $NN$ of
the $\Delta\Delta$ channel alone or of both $\Delta\Delta$ and $CC$ channels
brings a very small change in the $^3S_1$ and $^1S_0$ phase shifts below 2.5 fm
$^{-1}$, making the repulsion slightly weaker. The $CC$ channel brings slightly
more repulsion than the $\Delta\Delta$ channel. In fact the role of $CC$
channels is expected to increase for larger values of $k$, or alternatively
smaller separation distances between nucleons, where they could bring an
important contribution. Of course, the contribution of the $CC$ channels to
the $NN$ phase shifts vanishes at larger separations because of their color
structure. The conclusion regarding the minor contribution of $\Delta\Delta$
and $CC$ channels to the phase shifts below 2.5 fm$^{-1}$ is similar for
results  based on the OGE model (see for example \cite{FAE82}). Thus for $l=0$
waves it is good enough to perform one channel calculations in the lab energy
interval 0-350 MeV.

We recall that the pseudoscalar exchange interaction (\ref{GRAZ}) contains both
a short range part, responsible for the repulsion, and a long range Yukawa-type
potential which brings attraction in the $NN$ potential. In order to see the
difference in the amount of repulsion induced by the GBE and that induced by
the OGE interaction we repeated the one channel $(NN)$ calculations above by
removing the Yukawa-type part. We compared  the resulting phase shifts with
those of Fig. 2 of Ref. \cite{FAE82} obtained with an OGE interaction
parametrized such as to satisfy the stability condition (\ref{STABILITY}).
We found that in the GBE model the repulsion is much stronger and corresponds
to a hard core radius $r^{GBE}_0 = 0.68$ fm (versus $r^{OGE}_0 = 0.30$ fm) in
the $^3S_1$ and $r^{GBE}_0 = 0.81$ fm (versus $r^{OGE}_0 = 0.35$ fm) in the
$^1S_0$ partial waves. The radius $r_0$ was extracted from the phase shifts at
small $k$, which is approximately given by $\delta = -kr_0$. One can also see
that the repulsion induced by the GBE interaction in the $^3S_1$ partial wave
is weaker than that induced in the $^1S_0$ partial wave. This is consistent
with our previous result \cite{BAR99b} where we found that the height
of the repulsive core is lower for $^3S_1$ than for $^1S_0$, as mentioned in
the introduction. Thus the OGE model gives less replusion than the GBE model.
In Ref. \cite{SHI00} the stronger replusion induced by the GBE interaction is
viewed as a welcome feature in correctly describing the phase shifts above
$E_{lab}=350$ MeV.

A note of caution is required regarding the removal of the long-range Yukawa
part of the interaction (\ref{GRAZ}) with the parametrization (\ref{PARAM})
which contains a rather large coupling constant $g^2_{\eta^{'}q}/(4\pi)=2.7652$. The
$\eta^{'}$-meson exchange is responsible for describing correctly the
$\Delta- N$ splitting. If the long-range Yukawa part is removed, the model fails
to describe this splitting because the contribution coming from the
second term of (\ref{GRAZ}) for $\gamma = \eta^{'}$ becomes too large in a
$3q$ system in the parametrization (\ref{PARAM}).
We recall that the contribution to $N$ of the
short-range $\eta^{'}$-meson exchange part is proportional to a
factor of 2 and the contribution to $\Delta$ to a factor -2 \cite{GLO96a}, which brings
$\Delta$ too low and $N$ too high if the Yukawa part is removed. In these
circumstances two or three coupled channel calculations become meaningless.

It is also interesting to see the behaviour of the relative wave function
$\chi^{l=0}$ of Eq. (\ref{KAM5}) at short distances. Instead of $\chi^{l=0}$ it
is more appropriate \cite{OKA80} to introduce a renormalized wave function as
\begin{equation}\label{NORM}
\tilde{\chi}^{l=0}_{\alpha}(R) = \sum\limits_{\beta}\int{dR^{'}~[{N^{l=0}_{\beta\alpha}}(R,R')]^{1/2}
~\chi^{l=0}_{\beta}(R^{'})}\ ,
\end{equation}
where the quantity to be integrated contains the $l = 0$ component of the norm
$N$. In Fig. 3 we show results for the above function for the $^3S_1$ wave at
$k = 1$ fm$^{-1}$ both for the one and the three channel cases. One can see
that for $R<1$ fm the two functions are entirely different, in the three
channel case a node being present. If the renormalization was made with the
norm $N$ instead of its square, as in Eq. (\ref{NORM}), no node would have been
present. The existence of a node is related to the presence of the $[42]_{O}$
configuration in the wave function (see e.g. \cite{STA97}). Here, whenever it
appears, it is due to the cancellation of the positive and negative components
of the wave function, but the lack of a node does not exclude a repulsive
potential. In a renormalized wave function the amplitudes of positive and
negative components change their values depending on the multiplicative factor
$N$ or $N^{1/2}$ so the node could appear in one renormalization definition but
not in the other. On the other hand, as discussed above, the phase shift
changes insignificantly when one goes from one channel to three channels, and
this can also be seen in the asymptotic form of the wave function beyond $R=1$
fm, although in the overlap region the two functions are entirely different.
The above behaviour of the wave function is very similar to that found in Ref.
\cite{SHI00} where no long-range part is present in the schematic quark-quark
potential due to pion exchange. 

In Fig. 4 we represent the $^3S_1$ and $^1S_0$ phase shifts of Figs. 1 \& 2 in
the one channel case $(NN)$ again with the Yukawa part included, but this time
as a function of $E_{lab} = 2 \hbar^2 k^2/3m$ with $m = m_{u,d}$ of
(\ref{PARAM}). This is to show that in the GBE model the two phase shifts are
very near each other, with $\delta(^3S_1)$ slightly lower than $\delta(^1S_0)$.
Contrary, in OGE calculations as example those of Fig. 2 of Ref. \cite{FAE82}
one obtaines $\delta(^3S_1) > \delta(^1S_0)$. In calculations based on the OGE
model the difference between the two phase shifts is reduced by the addition of
a scalar potential acting at a nucleon level with a larger attractive strength
in the $^1S_0$ channel than in the $^3S_1$ channel  \cite{OKA83}. 
 
A major difference between the GBE $\delta(^3S_1)$ and $\delta(^1S_0)$ is
expected to appear after the inclusion of a quark-quark tensor force
\cite{PLE99}. This will modify only the $^3S_1$ phase shift.

\section{Conclusions}

This work is a further important step in our previous studies
\cite{BAR99a,BAR99b} of the $NN$ problem. We consider the two interacting
nucleons as a $6q$ system described by a Hamiltonian contining a linear
confinement plus a pseudoscalar (meson) exchange interaction between quarks.

Previously we derived an $NN$ potential in an adiabatic approximation. The
present study is based on a dynamical approach of the $NN$ interaction, namely
the resonating group method. We perform one, two and three coupled channel
calculations for the $^3S_1$ and $^1S_0$ phase shifts for laboratory energies
up to about 350 MeV.

Our conclusions are :

\begin{enumerate}

\item The phase shifts present a behaviour typical for strongly replusive
potentials. We find that this repulsion, which is induced by pseudoscalar
meson exchange is stronger than that produced by the OGE interaction.

\item In the $^1S_0$ partial wave the repulsion is stronger than in $^3S_1$
partial wave as our previous studies suggested.

\item Our results prove that in the laboratory energy interval 0-350 MeV the
one channel approximation is entirely satisfactory.

\end{enumerate}

Finally in future calculation, in order to describe the $^3S_1$ phase shift
the tensor force is compulsory and this is our following major step.

\vspace{2cm}
{\it Acknowledgements}. We are most grateful to Kiyotaka Shimizu for help in
understanding the resonating group method techniques and for constructive
criticism in preparing the manuscript.
\begin{center}
\appendix{\bf Appendix A}
\end{center}

The method to compute the six-body matrix elements is explained here using the
example of $S=1, I=0$ case.

We know that for the nucleon, the spin-flavor wavefunction is given by 
\begin{equation}
\psi_N=\frac{1}{\sqrt{2}}[\chi^{\rho}\phi^{\rho}+\chi^{\lambda}\phi^{\lambda}]\ ,
\end{equation}
where $\chi$ and $\phi$ are the spin and flavor parts respectively. For the
spin parts we have
\begin{eqnarray}\label{PSINCHI}
\chi^{\rho}_{1/2}&=&\frac{1}{\sqrt{2}}(\uparrow \downarrow \uparrow - \downarrow \uparrow \uparrow)\ ,\nonumber \\
\chi^{\rho}_{-1/2}&=&\frac{1}{\sqrt{2}}(\uparrow \downarrow \downarrow - \downarrow \uparrow \downarrow)\ ,\nonumber \\
\chi^{\lambda}_{1/2}&=&\frac{1}{\sqrt{6}}(\uparrow \downarrow \uparrow + \downarrow \uparrow \uparrow - 2 \uparrow \uparrow \downarrow)\ ,\nonumber \\
\chi^{\lambda}_{-1/2}&=&\frac{-1}{\sqrt{6}}(\uparrow \downarrow \downarrow + \downarrow \uparrow \downarrow -2 \downarrow \downarrow \uparrow )\ ,
\end{eqnarray}
and similarly for the flavor parts with $\uparrow$ replaced by $u$ and
$\downarrow$ replaced by $d$. Then for $\beta=NN$, the Eq. (\ref{PHI}) becomes
\begin{equation}\label{PHISI}
\Phi_{NN}^{SI}=\frac{1}{2}\sum C^{\frac{1}{2} \frac{1}{2} S}_{s_1 s_2 s}C^{\frac{1}{2} \frac{1}{2} I}_{\tau_1 \tau_2 \tau} [\chi^{\rho}_{s_1}(1)\phi^{\rho}_{\tau_1}(1)+\chi^{\lambda}_{s_1}(1)\phi^{\lambda}_{\tau_1}(1)][\chi^{\rho}_{s_2}(2)\phi^{\rho}_{\tau_2}(2)+\chi^{\lambda}_{s_2}(2)\phi^{\lambda}_{\tau_2}(2)]\ ,
\end{equation}
where $S$ and $I$ are the spin and isospin of the $NN$ system. $\chi(i)$ and $\phi(i)$ are the spin and flavor parts of the $i{\rm^{th}}$ nucleon. For 
$S=S_z=1$ and $I=I_z=0$, after inserting the values of the corresponding
Clebsch-Gordan coefficients we have
\begin{eqnarray}\label{PHISI10}
\Phi_{NN}^{10}&=&\frac{1}{2\sqrt{2}} \{[\chi^{\rho}_{1/2}(1)\phi^{\rho}_{1/2}(1)+\chi^{\lambda}_{1/2}(1)\phi^{\lambda}_{1/2}(1)][\chi^{\rho}_{1/2}(2)\phi^{\rho}_{-1/2}(2)+\chi^{\lambda}_{1/2}(2)\phi^{\lambda}_{-1/2}(2)]\nonumber \\
& & -[\chi^{\rho}_{1/2}(1)\phi^{\rho}_{-1/2}(1)+\chi^{\lambda}_{1/2}(1)\phi^{\lambda}_{-1/2}(1)][\chi^{\rho}_{1/2}(2)\phi^{\rho}_{1/2}(2)+\chi^{\lambda}_{1/2}(2)\phi^{\lambda}_{1/2}(2)] \}\ .
\end{eqnarray}
At this stage we use MATHEMATICA \cite{MATHEMATICA}. We introduce Eqs.
(\ref{PSINCHI}) and the equivalent for the flavor parts in (\ref{PHISI10}).
We get a huge expression with 338 terms depending now on the quantum numbers
of the quarks. In the matrix element of an operator $O$ we then get
$338^2=114244$ terms of the form
\begin{equation}
\langle s_1s_2s_3s_4s_5s_6\tau_1\tau_2\tau_3\tau_4\tau_5\tau_6|O|s_1's_2's_3's_4's_5's_6'\tau_1'\tau_2'\tau_3'\tau_4'\tau_5'\tau_6'\rangle\ ,
\end{equation}
where $s_i$ and $\tau_i$ ($i=1,\ldots,6$) stand for the spin and isospin
projection of the $i{\rm^{th}}$ quark. Note that the normal order of particles
is implied. Now let us choose 
$O = \vec{\sigma}_1\cdot\vec{\sigma}_3~\vec{\lambda}_1^f\cdot\vec{\lambda}_3^f
~P_{36}^{\sigma f}$, which contains the permutation P$_{36}$. Then we have
\begin{eqnarray}
&&\langle s_1s_2s_3s_4s_5s_6\tau_1\tau_2\tau_3\tau_4\tau_5\tau_6|\vec{\sigma}_1\cdot\vec{\sigma}_3\vec{\lambda}_1^f\cdot\vec{\lambda}_3^f P_{36}^{\sigma f}|s_1's_2's_3's_4's_5's_6'\tau_1'\tau_2'\tau_3'\tau_4'\tau_5'\tau_6'\rangle \nonumber \\
&&= \langle s_1s_2s_3s_4s_5s_6\tau_1\tau_2\tau_3\tau_4\tau_5\tau_6|\vec{\sigma}_1\cdot\vec{\sigma}_3\vec{\lambda}_1^f\cdot\vec{\lambda}_3^f|s_1's_2's_6's_4's_5's_3'\tau_1'\tau_2'\tau_6'\tau_4'\tau_5'\tau_3'\rangle\nonumber \\
&&= \langle s_1s_3\tau_1\tau_3|\vec{\sigma}_1\cdot\vec{\sigma}_3\vec{\lambda}_1^f\cdot\vec{\lambda}_3^f|s_1's_6'\tau_1'\tau_6'\rangle \ \delta_{s_2}^{s_2'}\delta_{s_4}^{s_4'}\delta_{s_5}^{s_5'}\delta_{s_6}^{s_3'}\delta_{\tau_2}^{\tau_2'}\delta_{\tau_4}^{\tau_4'}\delta_{\tau_5}^{\tau_5'}\delta_{\tau_6}^{\tau_3'}\nonumber \\
&&= \langle s_1s_3|\vec{\sigma}_1\cdot\vec{\sigma}_3|s_1's_6' \rangle \langle \tau_1\tau_3|\vec{\lambda}_1^f\cdot\vec{\lambda}_3^f|\tau_1'\tau_6' \rangle \ \delta_{s_2}^{s_2'}\delta_{s_4}^{s_4'}\delta_{s_5}^{s_5'}\delta_{s_6}^{s_3'}\delta_{\tau_2}^{\tau_2'}\delta_{\tau_4}^{\tau_4'}\delta_{\tau_5}^{\tau_5'}\delta_{\tau_6}^{\tau_3'}\ .
\end{eqnarray}
This shows how a six-body matrix element can be reduced to the calculation 
of two-body matrix elements. The necessary nonzero two-body matrix elements are
\begin{eqnarray}
\langle \uparrow \uparrow|\vec{\sigma}_1\cdot\vec{\sigma}_2|\uparrow \uparrow \rangle =\langle \downarrow \downarrow|\vec{\sigma}_1\cdot\vec{\sigma}_2|\downarrow \downarrow \rangle &=&1\ ,\nonumber \\
\langle \uparrow \downarrow|\vec{\sigma}_1\cdot\vec{\sigma}_2|\uparrow \downarrow \rangle =\langle \downarrow \uparrow|\vec{\sigma}_1\cdot\vec{\sigma}_2|\downarrow \uparrow \rangle &=&-1\ ,\nonumber \\
\langle \uparrow \downarrow|\vec{\sigma}_1\cdot\vec{\sigma}_2|\downarrow \uparrow \rangle =\langle \downarrow \uparrow|\vec{\sigma}_1\cdot\vec{\sigma}_2|\uparrow \downarrow \rangle &=&2\ ,\nonumber \\
\langle uu|\vec{\lambda}_1^f\cdot\vec{\lambda}_2^f|uu \rangle =\langle dd|\vec{\lambda}_1^f\cdot\vec{\lambda}_2^f|dd \rangle &=&4/3\ ,\nonumber \\
\langle ud|\vec{\lambda}_1^f\cdot\vec{\lambda}_2^f|ud \rangle =\langle du|\vec{\lambda}_1^f\cdot\vec{\lambda}_2^f|du \rangle &=&-2/3\ ,\nonumber \\
\langle ud|\vec{\lambda}_1^f\cdot\vec{\lambda}_2^f|du \rangle =\langle du|\vec{\lambda}_1^f\cdot\vec{\lambda}_2^f|ud \rangle &=&2\ .
\end{eqnarray}
MATHEMATICA is then used to compute systematically the sum of the 114244 terms
stemming from Eq. (\ref{PHISI10}).

In Tables I \& II all required six-body matrix elements obtained by this
technique are listed.


\begin{table}
\renewcommand{\arraystretch}{0.7}
\parbox{18cm}{\caption[matrixelements3]{\label{matrixelements3}  Matrix
elements $\langle \alpha|O|\beta \rangle$ of different operators $O$ for (S,I)
= (1,0).}}
\begin{tabular}{ccccccc}
$\alpha$ & $NN$ & $NN$  & $\Delta\Delta$ & $NN$ & $\Delta\Delta$ & $CC$ \\
$\beta$  & $NN$ & $\Delta\Delta$ & $\Delta\Delta$ & $CC$ & $CC$  & $CC$ \\
\tableline
$1$                                                                       
&   972 &    0 &   972 &    0 &    0 &   972 \\
$P_{36}^{f \sigma c}$                                                     
&   -12 &   48 &    12 & -144 &  288 &  -756 \\
$\lambda_1^c.\lambda_2^c$                                                 
& -2592 &    0 & -2592 &    0 &    0 &  -648 \\
$\lambda_3^c.\lambda_6^c$                                                 
&     0 &    0 &     0 &    0 &    0 & -1296 \\
$\lambda_1^c.\lambda_2^c\ P_{36}^{f \sigma c}$                            
&    32 & -128 &   -32 &  384 & -768 &    72 \\
$\lambda_3^c.\lambda_6^c\ P_{36}^{f \sigma c}$                            
&   -64 &  256 &    64 &   96 & -192 &  1152 \\
$\lambda_1^c.\lambda_3^c\ P_{36}^{f \sigma c}$                            
&    32 & -128 &   -32 &  384 & -768 &   720 \\
$\lambda_1^c.\lambda_6^c\ P_{36}^{f \sigma c}$                            
&    32 & -128 &   -32 &  -48 &   96 &   720 \\
$\lambda_1^c.\lambda_4^c\ P_{36}^{f \sigma c}$                            
&   -16 &   64 &    16 &   24 &  -48 &  1260 \\
$\sigma_1.\sigma_2\ \tau_1.\tau_2$                                        
&  4860 &    0 &   972 &    0 &    0 &   108 \\
$\sigma_3.\sigma_6\ \tau_3.\tau_6$                                        
&  -900 &  576 &  1980 &    0 &    0 &  1116 \\
$\sigma_1.\sigma_2\ \tau_1.\tau_2\ P_{36}^{f \sigma c}$                   
&  -444 &   48 &    12 & -720 &  288 &   588 \\
$\sigma_3.\sigma_6\ \tau_3.\tau_6\ P_{36}^{f \sigma c}$                   
&   708 &   48 &  1596 &  240 &  672 & -1092 \\
$\sigma_1.\sigma_3\ \tau_1.\tau_3\ P_{36}^{f \sigma c}$                   
&   132 &  336 &    12 & -720 &  288 &  -420 \\
$\sigma_1.\sigma_6\ \tau_1.\tau_6\ P_{36}^{f \sigma c}$                   
&   132 &   48 &    12 &  336 &  -96 &  -420 \\
$\sigma_1.\sigma_4\ \tau_1.\tau_4\ P_{36}^{f \sigma c}$                   
&    36 & -144 &   -36 &  228 &  288 & -1260 \\
$\sigma_1.\sigma_2\ \lambda_1^f.\lambda_2^f$                              
&  4536 &    0 &  1296 &    0 &    0 &   -18 \\
$\sigma_3.\sigma_6\ \lambda_3^f.\lambda_6^f$                              
&  -864 &  576 &  1584 &    0 &    0 &  1020 \\
$\sigma_1.\sigma_2\ \lambda_1^f.\lambda_2^f\ P_{36}^{f \sigma c}$         
&  -376 &   64 &    16 & -672 &  384 &   706 \\
$\sigma_3.\sigma_6\ \lambda_3^f.\lambda_6^f\ P_{36}^{f \sigma c}$         
&   784 &   32 &  1520 &  216 &  528 & -1024 \\
$\sigma_1.\sigma_3\ \lambda_1^f.\lambda_3^f\ P_{36}^{f \sigma c}$         
&   104 &  304 &    16 & -672 &  384 &  -332 \\
$\sigma_1.\sigma_6\ \lambda_1^f.\lambda_6^f\ P_{36}^{f \sigma c}$         
&   104 &   64 &    16 &  340 & -200 &  -332 \\
$\sigma_1.\sigma_4\ \lambda_1^f.\lambda_4^f\ P_{36}^{f \sigma c}$         
&    44 & -152 &   -32 &  278 &  164 & -1197 \\
$\sigma_1.\sigma_2\ \lambda_1^{f,0}.\lambda_2^{f,0}$                      
&  -648 &    0 &   648 &    0 &    0 &  -252 \\
$\sigma_3.\sigma_6\ \lambda_3^{f,0}.\lambda_6^{f,0}$                      
&    72 &    0 &  -792 &    0 &    0 &  -192 \\
$\sigma_1.\sigma_2\ \lambda_1^{f,0}..\lambda_2^{f,0}\ P_{36}^{f \sigma c}$ 
&   136 &   32 &     8 &   96 &  192 &   236 \\
$\sigma_3.\sigma_6\ \lambda_3^{f,0}.\lambda_6^{f,0}\ P_{36}^{f \sigma c}$ 
&   152 &  -32 &  -152 &  -48 & -288 &   136 \\
$\sigma_1.\sigma_3\ \lambda_1^{f,0}.\lambda_3^{f,0}\ P_{36}^{f \sigma c}$ 
&   -56 &  -64 &     8 &   96 &  192 &   176 \\
$\sigma_1.\sigma_6\ \lambda_1^{f,0}.\lambda_6^{f,0}\ P_{36}^{f \sigma c}$ 
&   -56 &   32 &     8 &    8 & -208 &   176 \\
$\sigma_1.\sigma_4\ \lambda_1^{f,0}.\lambda_4^{f,0}\ P_{36}^{f \sigma c}$ 
&    16 &  -16 &     8 &  -20 & -248 &   126 \\
\tableline
factor & $\frac{1}{972}$ & $\frac{\sqrt5}{972}$ & $\frac{1}{972}$ 
& $\frac{\sqrt5}{972}$ & $\frac{1}{972}$ & $\frac{1}{972}$ \\
\end{tabular}
\end{table}

\begin{table}
\renewcommand{\arraystretch}{0.7}
\parbox{18cm}{\caption[matrixelements4]{\label{matrixelements4}  Matrix
elements $\langle \alpha|O|\beta \rangle$ of different operators $O$ for (S,I)
= (0,1).}}
\begin{tabular}{ccccccc}
$\alpha$ & $NN$ & $NN$ & $\Delta\Delta$ & $NN$ & $\Delta\Delta$ & $CC$ \\
$\beta$  & $NN$ & $\Delta\Delta$ & $\Delta\Delta$ & $CC$ & $CC$ & $CC$ \\
\tableline
$1$                                                                       
&   972 &    0 &   972 &    0 &    0 &   972 \\
$P_{36}^{f \sigma c}$                                                     
&   -12 &   48 &    12 & -144 &  288 &  -756 \\
$\lambda_1^c.\lambda_2^c$                                                 
& -2592 &    0 & -2592 &    0 &    0 &  -648 \\
$\lambda_3^c.\lambda_6^c$                                                 
&     0 &    0 &     0 &    0 &    0 & -1296 \\
$\lambda_1^c.\lambda_2^c\ P_{36}^{f \sigma c}$                            
&    32 & -128 &   -32 &  384 & -768 &    72 \\
$\lambda_3^c.\lambda_6^c\ P_{36}^{f \sigma c}$                            
&   -64 &  256 &    64 &   96 & -192 &  1152 \\
$\lambda_1^c.\lambda_3^c\ P_{36}^{f \sigma c}$                            
&    32 & -128 &   -32 &  384 & -768 &   720 \\
$\lambda_1^c.\lambda_6^c\ P_{36}^{f \sigma c}$                            
&    32 & -128 &   -32 &  -48 &   96 &   720 \\
$\lambda_1^c.\lambda_4^c\ P_{36}^{f \sigma c}$                            
&   -16 &   64 &    16 &   24 &  -48 &  1260 \\
$\sigma_1.\sigma_2\ \tau_1.\tau_2$                                        
&  4860 &    0 &   972 &    0 &    0 &   108 \\
$\sigma_3.\sigma_6\ \tau_3.\tau_6$                                        
&  -900 &  576 &  1980 &    0 &    0 &  1116 \\
$\sigma_1.\sigma_2\ \tau_1.\tau_2\ P_{36}^{f \sigma c}$                   
&  -444 &   48 &    12 & -720 &  288 &   588 \\
$\sigma_3.\sigma_6\ \tau_3.\tau_6\ P_{36}^{f \sigma c}$                   
&   708 &   48 &  1596 &  240 &  672 & -1092 \\
$\sigma_1.\sigma_3\ \tau_1.\tau_3\ P_{36}^{f \sigma c}$                   
&   132 &  336 &    12 & -720 &  288 &  -420 \\
$\sigma_1.\sigma_6\ \tau_1.\tau_6\ P_{36}^{f \sigma c}$                   
&   132 &   48 &    12 &  336 &  -96 &  -420 \\
$\sigma_1.\sigma_4\ \tau_1.\tau_4\ P_{36}^{f \sigma c}$                   
&    36 & -144 &   -36 &  228 &  288 & -1260 \\
$\sigma_1.\sigma_2\ \lambda_1^f.\lambda_2^f$                              
&  4536 &    0 &  1296 &    0 &    0 &  -126 \\
$\sigma_3.\sigma_6\ \lambda_3^f.\lambda_6^f$                              
& -1008 &  576 &  1440 &    0 &    0 &   948 \\
$\sigma_1.\sigma_2\ \lambda_1^f.\lambda_2^f\ P_{36}^{f \sigma c}$         
&  -376 &   64 &    16 & -672 &  384 &   814 \\
$\sigma_3.\sigma_6\ \lambda_3^f.\lambda_6^f\ P_{36}^{f \sigma c}$         
&   832 &   32 &  1568 &  232 &  496 &  -976 \\
$\sigma_1.\sigma_3\ \lambda_1^f.\lambda_3^f\ P_{36}^{f \sigma c}$         
&   104 &  304 &    16 & -672 &  384 &  -260 \\
$\sigma_1.\sigma_6\ \lambda_1^f.\lambda_6^f\ P_{36}^{f \sigma c}$         
&   104 &   64 &    16 &  364 & -248 &  -260 \\
$\sigma_1.\sigma_4\ \lambda_1^f.\lambda_4^f\ P_{36}^{f \sigma c}$         
&    36 & -168 &   -48 &  298 &  124 & -1155 \\
$\sigma_1.\sigma_2\ \lambda_1^{f,0}.\lambda_2^{f,0}$                      
&  -648 &    0 &   648 &    0 &    0 &  -468 \\
$\sigma_3.\sigma_6\ \lambda_3^{f,0}.\lambda_6^{f,0}$                      
&  -216 &    0 & -1080 &    0 &    0 &  -336 \\
$\sigma_1.\sigma_2\ \lambda_1^{f,0}.\lambda_2^{f,0}\ P_{36}^{f \sigma c}$ 
&   136 &   32 &     8 &   96 &  192 &   452 \\
$\sigma_3.\sigma_6\ \lambda_3^{f,0}.\lambda_6^{f,0}\ P_{36}^{f \sigma c}$ 
&   248 &  -32 &   -56 &  -16 & -352 &   232 \\
$\sigma_1.\sigma_3\ \lambda_1^{f,0}.\lambda_3^{f,0}\ P_{36}^{f \sigma c}$ 
&   -56 &  -64 &     8 &   96 &  192 &   320 \\
$\sigma_1.\sigma_6\ \lambda_1^{f,0}.\lambda_6^{f,0}\ P_{36}^{f \sigma c}$ 
&   -56 &   32 &     8 &   56 & -304 &   320 \\
$\sigma_1.\sigma_4\ \lambda_1^{f,0}.\lambda_4^{f,0}\ P_{36}^{f \sigma c}$ 
&     0 &  -48 &   -24 &   20 & -328 &   210 \\
\tableline
factor & $\frac{1}{972}$ & $\frac{\sqrt5}{972}$ & $\frac{1}{972}$ 
& $\frac{\sqrt5}{972}$ & $\frac{1}{972}$ & $\frac{1}{972}$ \\
\end{tabular}
\end{table}

\begin{figure}
\begin{center}
\psfig{figure=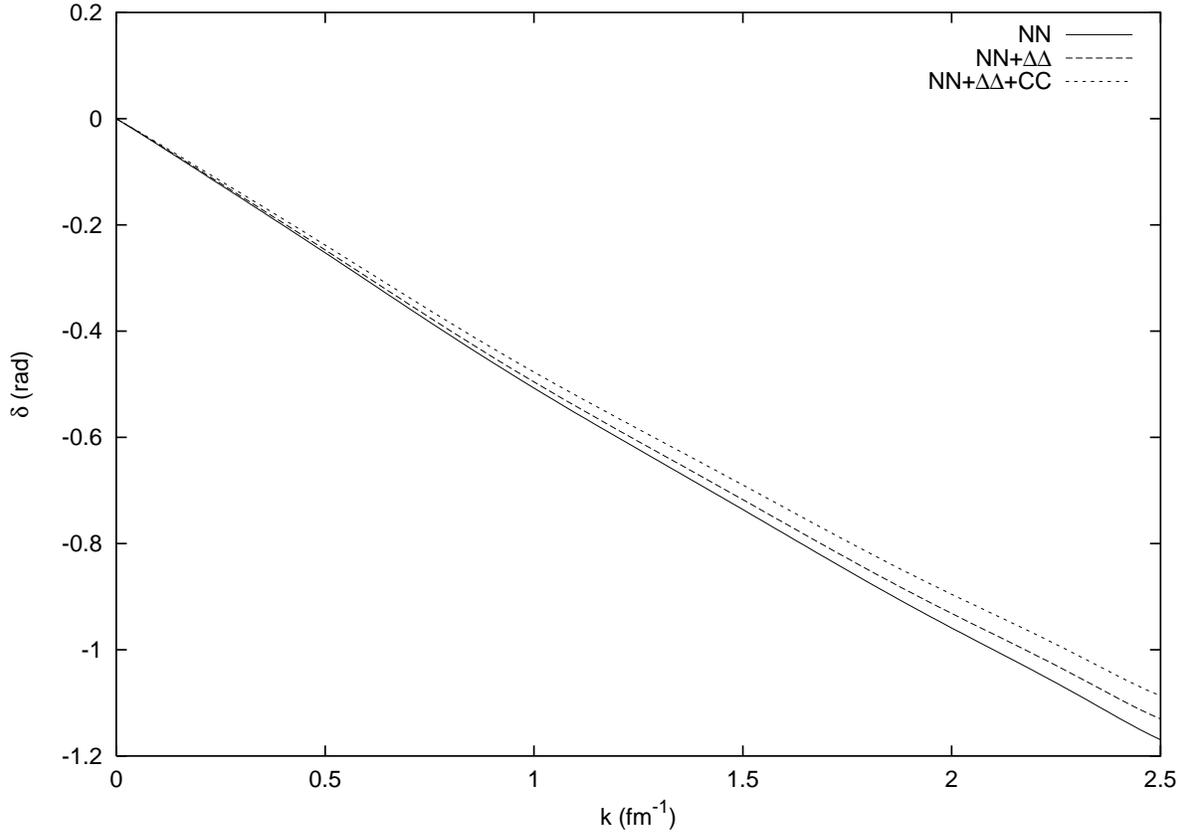,width=16cm}
\end{center}
\caption{\label{Fig. 1} $^3S_1$ NN scattering phase shift as a function of $k$.
The solid line shows the result for the $NN$ channel only, the dotted line for
the $NN$+$\Delta\Delta$  and the dashed line for the $NN$+$\Delta\Delta$+$CC$
coupled channels.}
\end{figure}

\begin{figure}
\begin{center}
\psfig{figure=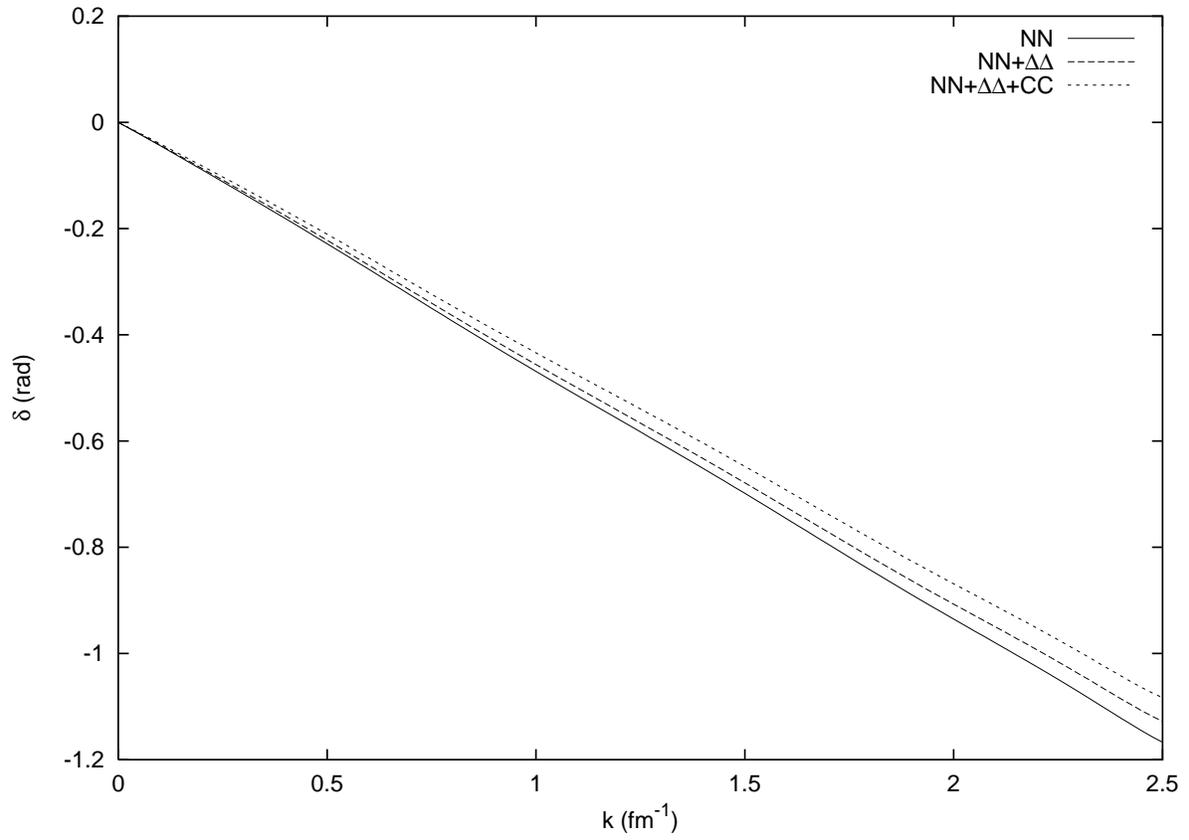,width=16cm}
\end{center}
\caption{\label{Fig. 2}  Same as Fig. 1 but for the $^1S_0$ partial wave.}
\end{figure}
 
\begin{figure}
\begin{center}
\psfig{figure=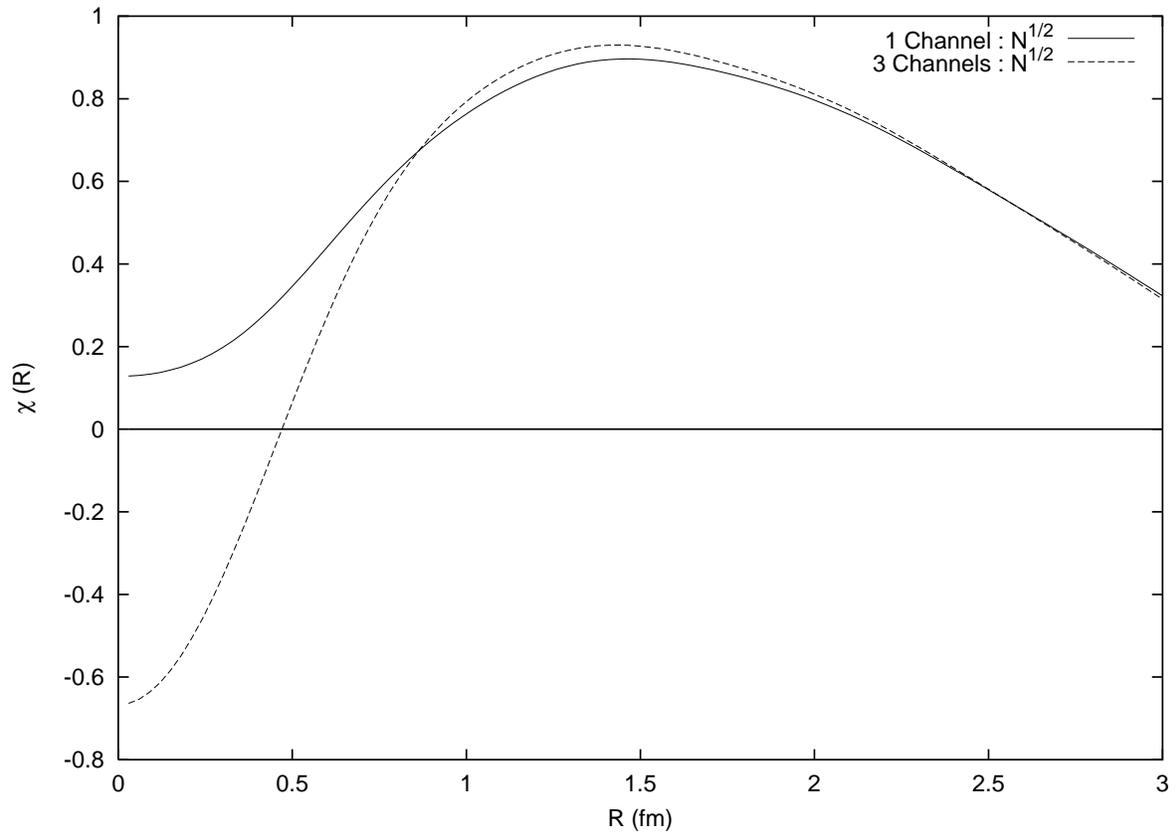,width=16cm}
\end{center}
\caption{\label{Fig. 3} The relative wave function of Eq. (\ref{NORM}) for the
$^3S_1$ partial wave for $k = 1$ fm$^{-1}$ obtained in one channel (solid line)
and three channels (dashed line) calculations.}
\end{figure}

\begin{figure}
\begin{center}
\psfig{figure=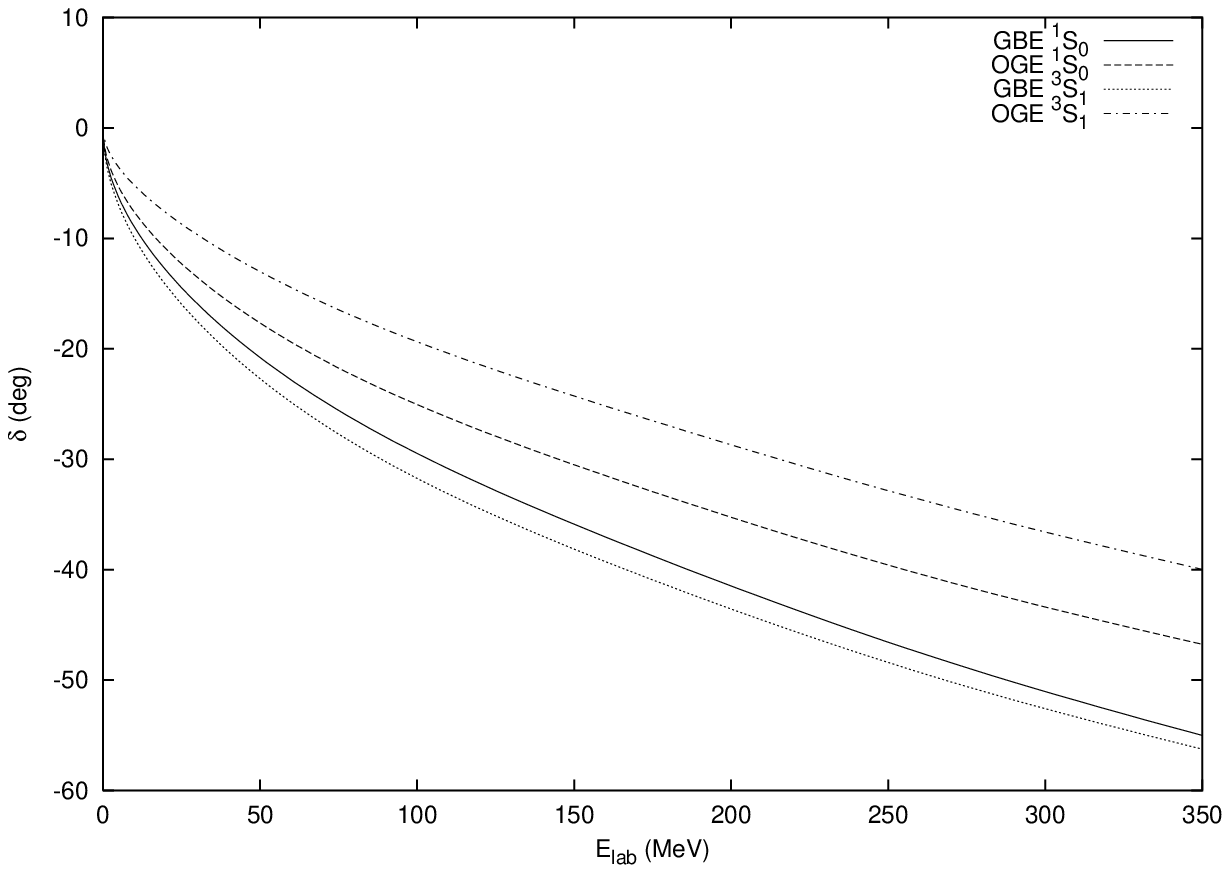,width=16cm}
\end{center}
\caption{\label{Fig. 4} $^1S_0$ and $^3S_1$ $NN$ scattering phase shifts as a
function of the laboratory energy $E_{lab}$. The solid and dotted lines show
the result corresponding to the GBE model and the dashed and dot-dashed lines
that of the OGE model (see Ref.\protect\cite{FAE82}). }
\end{figure}


\begin{references}

\bibitem{OKA85}
{M. Oka and K. Yazaki, ``Quarks and Nuclei", International Review of Nuclear
Physics, vol. 1, ed. W. Weise (World Scientific, Singapore, 1985), p. 489}

\bibitem{MYH88}
{F. Myhrer and J. Wroldsen, Rev. Mod. Phys. {\bf 60} (1988) 629}

\bibitem{SHI89}
{K. Shimizu, Rep. Prog. Phys. {\bf 52} (1989) 1 and references therein}

\bibitem{STB00}
{K. Shimizu, S. Tacheuchi and A. J. Buchmann, Progr. Theor. Phys. Suppl.
{\bf 137} (2000) 43}

\bibitem{OKA83}
{M. Oka and K. Yazaki, Nucl. Phys. {\bf A402} (1983) 447}

\bibitem{KUS91}
{A. M. Kusainov, V. G. Neudatchin and I. T. Obukhovsky, Phys. Rev. {\bf C44} (1991)
2343}

\bibitem{ZHA94}
{Z. Zhang, A. Faessler, U. Straub and L. Ya. Glozman, Nucl. Phys. {\bf A578}
(1994) 573; A. Valcarce, A. Buchmann, F. Fern\'andez and A. Faessler,
Phys. Rev. {\bf C50} (1994) 2246}

\bibitem{FUJ96}
{Y. Fujiwara, C. Nakamoto and Y. Suzuki, Phys. Rev. Lett. {\bf 76} (1996) 2242}

\bibitem{GLO96a}
{L. Ya. Glozman and D. O. Riska, Phys. Rep. {\bf 268} (1996) 263}

\bibitem{GLO96b}
{L. Ya. Glozman, Z. Papp and W. Plessas, Phys. Lett. {\bf B381}
(1996) 311}

\bibitem{GLO97}
{L. Ya. Glozman, Z. Papp, W. Plessas, K. Varga and R. F. Wagenbrunn, Nucl. Phys.
{\bf A623} (1997) 90c}

\bibitem{GLO98}
{L. Ya. Glozman, W. Plessas, K. Varga and R. F. Wagenbrunn, Phys. Rev.
{\bf D58} (1998) 094030}

\bibitem{GLO00}
{L. Ya. Glozman, Nucl. Phys. {\bf A663\&664} (2000) 103c}

\bibitem{MAN94}
{A. Manohar and H. Georgi, Nucl. Phys. {\bf B234} (1994) 189}

\bibitem{ITW88}
{It was noticed first by D. Robson, see {\it Proc. Topical Conf. on Nuclear 
Chromodynamics}, World Scientific (Singapore 1988) that a flavor-spin
interaction of type $ - \sum_{i<j}  \lambda_{i}^{F} . \lambda_{j}^{F}
\vec{\sigma}_i . \vec{\sigma}_j$ gives the position of the Roper resonance
below the first excited negative parity states, in agreement with the
experiment. However Robson's conclusion that any contact interaction,
irrespective of its flavor-spin or color-spin structure, would give the
desired order, was incorrect}  

\bibitem{RIS99}
{D. O. Riska and G. E. Brown, Nucl. Phys. {\bf A653} (1999) 251}

\bibitem{COL99}
{H. Collins and H. Georgi, Phys. Rev. {\bf D59} (1999) 094010}

\bibitem{CAR00}
{C. D. Carone, Nucl. Phys. {\bf A663\&664} (2000) 687c}

\bibitem{SAS00}
{S. Sasaki, hep-ph/0004252}

\bibitem{STA97}
{Fl. Stancu, S. Pepin and L. Ya. Glozman, Phys. Rev. {\bf C56} (1997) 2779}

\bibitem{BAR99a}
{D. Bartz and Fl. Stancu, Phys. Rev. {\bf C59} (1999) 1756}

\bibitem{BAR99b}
{D. Bartz and Fl. Stancu, Phys. Rev. {\bf C60} (1999) 055207}

\bibitem{WHE37}
{J. A. Wheeler, Phys. Rev. {\bf 52} (1937), 1107}

\bibitem{WIL77}
{K. Wildermuth and Y. C. Tang, {\it A Unified Theory of the Nucleus} (Vieweg
Braunschweig, 1977)}

\bibitem{KAM77}
{M. Kamimura, Prog. Theor. Phys. Suppl. {\bf 62} (1977) 236}

\bibitem{STA99}
{Fl. Stancu and L. Ya. Glozman, nucl-th/9906058}

\bibitem{OKA80}
{M. Oka and K. Yazaki, Phys. Lett. {\bf 90B} (1980) 41, M. Oka and K. Yazaki, 
Progr. Theor. Phys. {\bf 66} (1981) 556, 572}

\bibitem{CVE83}
{M. Cveti\v{c},B. Golli, N. Manko\v{c}-Bor\v{s}tnik and M. Rosina,
Nucl. Phys. {\bf A395} (1983) 349}

\bibitem{ABR64}
{M. Abramowitz and I. A. Stegun, {\it Handbook of Mathematical Functions}
(Dover edition, 1964)}

\bibitem{HAR81}
{M. Harvey, Nucl. Phys. {\bf A352} (1981) 301 and 326; {\bf A481} (1988) 834}

\bibitem{FAE82}
{A. Faessler, F. Fern\'andez, G. L\"ubeck and K. Shimizu, Nucl. Phys.
{\bf A402} (1983) 555}

\bibitem{KS84}
{K. Shimizu, Phys. Let. {\bf 148B} (1984) 418}

\bibitem{SHI00}
{K. Shimizu and L. Ya. Glozman, Phys. Lett. {\bf B477} (2000) 59}

\bibitem{PLE99}
{W. Plessas, L. Ya. Glozman, K. Varga and R. F. Wagenbrunn, Proc. 2nd
International Conference on {\it Perspectives in Hadronic Physics}, Trieste,
1999, eds. S. Boffi et al., (World Scientific, Singapore, 2000), p. 136;
see also R. F. Wagenbrunn, L. Ya. Glozman, W. Plessas, K. Varga, Nucl. Phys.
{\bf A663\&664} (2000) 703c}

\bibitem{MATHEMATICA}
{S. Wolfram, The Mathematica book, Wolfram Media/Cambridge University Press,
Cambridge, 1996}

\end{references}
\end{document}